\documentclass[aps,prd,preprint,tightenlines,superscriptaddress,
amsfonts,amssymb,amsmath,byrevtex,showpacs,nofootinbib]{revtex4-2}
\usepackage[polish,english]{babel}
\usepackage[utf8]{inputenc}
\usepackage[T1]{fontenc}
\usepackage{ulem}
\usepackage{latexsym}
\usepackage{graphicx}
\usepackage{geometry}
\usepackage{epstopdf}
\usepackage{subfig}
\usepackage{color}

\graphicspath{ {./}{./rys/} } 
\usepackage{xcolor}
\usepackage{pdfsync}
\usepackage{fancyhdr}
\usepackage{makeidx}
\usepackage[breaklinks,colorlinks,backref]{hyperref}
\hypersetup{
    colorlinks, 
    linktoc=all, 
    linkcolor=red,  
  citecolor=hyptxt,
  urlcolor=blue}
  \hypersetup{
  citebordercolor=,
  filebordercolor=green,
  linkbordercolor=blue
}
\definecolor{hyptxt}{rgb}{0.7, 0.4, 0.9}
\let\babellll\lll
\let\lll\relax
\usepackage{amsmath}
\let\lll\babellll 
\usepackage{tensor}

\newcommand{\RNumb}{\mathbb{R}}
\newcommand{\CNumb}{\mathbb{C}}

\newcommand{\UnitOp}{\hat{1\kern-4.75pt 1}} 
\newcommand{\MatUnit}{1\kern-3pt 1} 
\newcommand{\Trace}{\mathrm{Tr}} 
\newcommand{\Prob}{\mathrm{Prob}\,} 
\newcommand{\Group}[1]{\textrm{#1}} 

\newcommand{\HW}[1]{\mathcal{HW}(#1)} 
\newcommand{\Bra}[1]{\langle #1 \vert} 
\newcommand{\Ket}[1]{\vert #1 \rangle} 
\newcommand{\BraKet}[2]{\langle #1 \vert #2 \rangle} 
\newcommand{\Aver}[1]{\langle #1 \rangle} 
\newcommand{\Norm}[1]{\|\kern.3ex#1\kern.3ex \|} 

\newcommand{\Var}[1]{\mathrm{var}(#1)} 
\newcommand{\EOp}{\mathsf{E}\kern-1pt\llap{$\vert$}}   
\newcommand{\WOp}{\hat{\mathsf{W}\kern-1pt\llap{$-$}}} 
\newcommand{\FOp}{\mathsf{F}\kern-1pt\llap{$\vert$}}   

\newcommand{\StateSpace}[1]{\mathcal #1} 
\newcommand{\Komentarz}[1]{} 

\usepackage{accents}

\begin{document}

\title{Integral quantization based on the Heisenberg-Weyl group}

\author{Aleksandra P\c{e}drak} \email{aleksandra.pedrak@ncbj.gov.pl}
\affiliation{Department of Fundamental Research, National Centre for Nuclear
  Research, Pasteura 7, 02-093 Warszawa, Poland}

\author{Andrzej G\'{o}\'{z}d\'{z}}
\email{andrzej.gozdz@umcs.lublin.pl}
\affiliation{Institute of Physics, Maria Curie-Sk{\l}odowska
University, pl.\  Marii Curie-Sk{\l}odowskiej 1, 20-031 Lublin, Poland}

\author{W{\l}odzimierz Piechocki} \email{wlodzimierz.piechocki@ncbj.gov.pl}
\affiliation{Department of Fundamental Research, National Centre for Nuclear
  Research, Pasteura 7, 02-093 Warszawa, Poland}

\author{Patryk Mach} \email{patryk.mach@uj.edu.pl} \affiliation{Institute of
  Theoretical Physics, Jagiellonian University in Krak\'{o}w,
  {\L}ojasiewicza 11, 30-348 Krak\'{o}w, Poland}

\author{Adam Cie\'{s}lik} \email{adam.cieslik@doctoral.uj.edu.pl}
\affiliation{Institute of Theoretical Physics, Jagiellonian University in
  Krak\'{o}w, {\L}ojasiewicza 11, 30-348 Krak\'{o}w, Poland}

\date{\today}

\begin{abstract}
We develop a relativistic framework of integral quantization applied to the
motion of spinless particles in the four-dimensional Minkowski spacetime. The
proposed scheme is based on coherent states generated by the action of the
Heisenberg-Weyl group and has been motivated by the Hamiltonian description of
the geodesic motion in General Relativity. We believe that this formulation
should also allow for a generalization to the motion of test particles in curved
spacetimes. A key element in our construction is the use of suitably defined
positive operator-valued measures. We show that this approach can be used to
quantize the one-dimensional nonrelativistic harmonic oscillator, recovering the standard
Hamiltonian as obtained by the canonical quantization. A direct application of our model,
including a computation of transition amplitudes between states characterized by fixed
positions and momenta, is postponed to a forthcoming article.
\end{abstract}


\maketitle

\tableofcontents

\section{Introduction}\label{Int}

This work presents a generalization of the so-called integral quantization
(IQ) method that is a special case of quantization based on deformations of
quantum measures. The IQ has been used in quantizations of
numerous physical systems. For a comprehensive review of applications we
recommend references \cite{JPG1,Klauder,Proc, JPG2}.
We have already used the IQ approach based on the affine group. See, for
instance, \cite{ast1,ast2,ast3,ast4} for applications in astrophysics and
cosmology \cite{cos1,cos2,cos3,met}.
In articles  \cite{hyb1,hyb2,hyb3,hyb4} a hybrid
of affine and canonical quantizations was used.

The IQ method turns out to be  highly efficient in ascribing a quantum system to a given
classical one. Its important feature is the ability of reproducing classical observables
via expectation values of corresponding quantized  observables. It is also a mathematically well-defined
procedure. As an ultimate goal, we believe that our method may contribute to the construction of a theory of quantum gravity.

In this work we apply the IQ to the motion of a relativistic spinless particle
in the Minkowski spacetime. We use the Hamiltonian description of the geodesic
motion, known from General Relativity, and construct a scheme which, in
principle, could be also applied to quantize the motion of a free particle in a
curved spacetime.  In the context of the motion in the Minkowski spacetime, this
Hamiltonian approach suggests to construct the IQ procedure basing on the
Heisenberg-Weyl group, as it allows for a one-to-one correspondence between the
positions and momenta of the classical phase space and the group parameters.

Traditionally, in quantum mechanics time is not considered to be a physical degree of freedom of a system,
but rather a parameter. The IQ method enables a framework in which the time can be treated as a quantum observable.
Such an approach is supported by series of experiments (see, e.g., \cite{PEv2023,PEvClock2024,third,Hor} and references therein).
They seem to imply that in the quantum
world treating the time as a parameter is only an approximation, although in most cases an excellent one.
In addition, promoting the time to an observable makes the quantization procedure
more unique.

Attempts to treat time as a physical degree of freedom are (to various extents) usually made in the general
context of quantum gravity. This and related issues are sometimes referred broadly as the time problem in quantum
gravity (see, e.g., \cite{Isham} or \cite{Maniccia} for a recent review).
Notable approaches include the Wentzel–Kramer–Brillouin (WKB) approximation to the Wheeler-DeWitt equation (see \cite{Maniccia} and references therein) or
works by Kuchar and Torre \cite{KucharTorre} or Brown and Kuchar \cite{BrownKuchar}, and many others.
In a similar spirit, Hartle and Kuchar \cite{Hartle} were able to rewrite the path-integral canonical
quantization of a relativistic free spinless particle using covariant four dimensional integration measures.
The starting point in this approach consists in reparametrizing the time variable so that the resulting scheme
can serve as a model of theories in which time is treated as one of dynamical variables.

In this paper we describe the main elements of our model, focusing on formal
details of the IQ. In particular, we pay special attention to a formulation in
terms of the so-called positive operator valued measures (POVM), which leads to
uniqueness in ascribing quantum operators to classical observables (see, e.g.,
\cite{Busch1996,Gazeau2015} and references therein).  We introduce the
eigenstates of the position and momentum operators and solve the eigenvalue
problem of the quantum Hamiltonian associated with classical geodesics. As a
test of the IQ based on the Heisenberg-Weyl group, we show that this method
allows one to recover the results obtained by the canonical quantization for the
standard nonrelativistic one-dimensional harmonic oscillator problem.

Sample applications of the formalism introduced in this paper are discussed in \cite{QMin}. They include a computation of transition amplitudes between two coherent states. Such amplitudes allow us to model qunatum random walks in the Minkowski spacetime and to recover interference patters in the standard double-slit experiment. The presented model seems to allow future generalizations to quantization of geodesic motion in curved spacetimes, including black-hole spacetimes such as Schwarzschild and
Kerr. This might have observational consequences, for instance in the context of shadows of black holes.

The paper is organized as follows: After specifying conventions concerning the
Minkowski spacetime in Sec.\,II, we present the IQ based on the
Heisenberg-Weyl group in Sec.\,III. The POVM approach is presented in Sec.\,IV
with some applications. Sec.\,V concerns relativistic models. We conclude in Sec.\,VI.

Throughout the paper we use geometric units with $c = 1$, where $c$ denotes
the speed of light. We assume the metric signature $(-,+,+,+)$.

\section{Preliminaries}\label{Pre}

Geodesic equations, describing the motion of free test particles in General
Relativity, can be written in the Hamiltonian form
\begin{equation}
\frac{d x^\mu}{d \tilde s} =
\frac{\partial H}{\partial p_\mu}, \quad \frac{d p_\mu}{d \tilde s}
= - \frac{\partial H}{\partial x^\mu}.
\end{equation}
Here $x^\mu$, $\mu = 0, 1, 2, 3$, denote the coordinates along the geodesic. The
corresponding covariant momentum components are defined as $p_\mu = g_{\mu \nu}
p^\nu = g_{\mu \nu} d x^\nu / d \tilde s$. The Hamiltonian $H$, associated with
with the geodesic equations, can be written as
\begin{equation}
\label{gHam}
H = \frac{1}{2} g^{\mu\nu}(x) p_{\mu} p_{\nu},
\end{equation}
where $g^{\mu \nu}$ denote the contravariant components of the metric
tensor. The affine parameter $\tilde s$ can be chosen in such a way that
$g^{\mu\nu} p_\mu p_\nu = -m^2$, where $m$ is the rest mass of the particle
moving along the geodesic. Consequently $H = - \frac{1}{2}m^2$. For timelike
geodesics, the proper time $s$ is related to the affine parameter $\tilde s$ by
$\tilde s = s/m$.

We propose a semiclassical quantization scheme which, in principle, can be
generalized to a quantization of the geodesic motion in curved spacetimes,
described by the geodesic Hamiltonian (\ref{gHam}). In this paper we deal with
the simplest, yet challenging, example of the motion in the flat Minkowski
spacetime. In a sense, this renders the Hamiltonian description unnecessary
(geodesics are simply straight lines in the Minkowski spacetime), but we adhere
to this formalism to allow for future generalizations and as a guideline for the
quantization procedure.

In the case of Minkowski spacetime the metric does not depend on $x$ and is
represented by the diagonal matrix $g=\mathrm{(-1,+1,+1,+1)}$ so that we have
(in Cartesian coordinates associated with an orthogonal frame)
$g_{\mu\nu} x^\mu x^\nu = - x_0^2 + x_1^2 + x_2^2 + x_3^2$ and
$g^{\mu\nu} p_\mu p_\nu = - p_0^2 + p_1^2 + p_2^2 + p_3^2$.  The components of
the four-momentum satisfy
$-p_0 = p^0 = dx^0/d \tilde s,~p_i = p^i = dx^i/d \tilde s$, $i = 1,2,3$.

The variables $p_\mu$ and $x^\nu$ are independent and they define the phase
space
$\{(p_\mu,x^\nu) \colon \mu, \nu = 0,1,2,3 \} \cong \RNumb^4 \times \RNumb^4$,
which can be identified with the cotangent bundle $T^\star \mathcal{M}$ of the
Minkowski spacetime $(\mathcal{M}, g)$.

\section{Integral quantization}\label{Integral}

The IQ procedure requires a specification of the group $\Group{G}$ that can be
ascribed uniquely to the classical phase space of a given system. In this paper
we choose the Heisenberg-Weyl group $\HW{4}$ to play that role, as this group
can be identified with the cotangent bundle $T^\star \mathcal{M}$ of the Minkowski
spacetime.  The IQ based on the Heisenberg-Weyl group has yet another
advantage---the results obtained by this procedure remain, in many cases,
consistent with the outcomes of the canonical quantization.

The group $\HW{4}$ has a unitary irreducible representation in the carrier
Hilbert space $\StateSpace{K}=L^2(\RNumb^4,d^4\xi)$, consisting of square
integrable complex functions of four real variables $\xi^\mu$, which enables us
to construct the set of so-called coherent states in $\StateSpace{K}$.

\subsection{The Heisenberg-Weyl group in four dimensions}

Elements of the Heiseberg-Weyl group $\HW{4}$ are defined by 9 independent
generators: four coordinate operators $\hat{Q}^\mu$, four momentum operators
$\hat{P}_\mu$, and the unit operator $\UnitOp$, which satisfy the following
commutation relations
\begin{eqnarray}
&& [\hat{Q}^\mu,\hat{P}_\nu]=i\hbar \delta^\mu_\nu \UnitOp \, ,\\
&& [\hat{Q}^\mu,\UnitOp]=0 \, ,\\
&& [\hat{P}^\mu,\UnitOp]=0 \, ,\\
&& [\hat{Q}^\mu,\hat{Q}^\nu]=0 \, ,\\
&& [\hat{P}_\mu,\hat{P}_\nu]=0 \, ,
\end{eqnarray}
where $\delta^\mu_\nu$ denotes the Kronecker delta, and $\mu, \nu = 0,1,2,3$.
The required realization of the above commutation relations in the space
$\StateSpace{K}$ can be defined by the following action of the operators $\hat{Q}^\mu$
and $\hat{P}_\mu$:
\begin{eqnarray}
&&\hat{Q}^\mu\psi(\xi)=\xi^\mu \psi(\xi) \, , \label{QK} \\
&&\hat{P}_\mu\psi(\xi)
=-i\hbar\frac{\partial}{\partial \xi^\mu}\psi(\xi) \,. \label{PK}
\end{eqnarray}
Every element of the $\HW{4}$ group can be written as the
following unitary operator in $\StateSpace{K}$
\begin{equation}
\label{HW4Oper}
g(\kappa;p,x) = g(\kappa;p_0,\dots, p_{3},x^0,\dots,x^{3})
=\exp\left(i\kappa\UnitOp
+\frac{i}{\hbar}(p_\mu \hat{Q}^\mu-x^\mu \hat{P}_\mu)\right)
\, ,
\end{equation}
where $x^\mu$, $p_\mu$, and $\kappa$ denote group parameters.
The multiplication law for the group reads\footnote{We use the identity
\begin{equation}
\exp A\exp B=\exp\left(\frac{1}{2}[A,B]\right)\exp(A+B)\nonumber \, ,
\end{equation}
which is valid if
\begin{equation}
[A,[A,B]]=0,\qquad [B,[A,B]]=0 \,\nonumber .
\end{equation}
}
\begin{equation}
\label{HW4MultipLaw}
g(\kappa;p,x)g(\kappa'; p',x')=
g\left(\kappa+\kappa'-\frac{1}{2\hbar}(p'_\mu x^\mu -p_\mu x'^\mu );\;p+p',\;
x+x'\right)  \, .
\end{equation}
The Haar measure associated with the group $\HW{4}$ has the form $d\mu(\kappa,p,x):=
d\kappa\, d\rho(p,x)$, where
\begin{equation}
\label{HW4VolElem}
d\rho(p,x)= d^{4}p\, d^{4}x
:= dp_0\,dp_1dp_{2} dp_{3}\, dx^0\, dx^1 dx^{2} dx^{3} \, .
\end{equation}
The action of unitary operators \eqref{HW4Oper} in $\StateSpace{K}$
is given by
\begin{equation}
\hat{\mathcal{U}}(\kappa;\,p,x)\psi(\xi)
=\exp(i\kappa)\exp\left(\frac{-ip_\mu x^\mu}{2\hbar}\right)
\exp\left(\frac{ip_\mu \xi^\mu}{\hbar}\right)\psi(\xi-x) \, ,
\end{equation}
where we change the notation from $g(\kappa;p,x)$ to
$\hat{\mathcal{U}}(\kappa;\,p,x)$ to emphasise that in the latter case the
action of operators $\hat Q^\mu$ and $\hat P_\mu$ is defined by Eqs.\ (\ref{QK})
and (\ref{PK}).

The subgroup parameterized by $\kappa$ as
\begin{equation}
g(\kappa;0,0)=\exp(i\kappa)\UnitOp \, ,
\end{equation}
forms the unitary group $\Group{U(1)}$, which is the center of the
Heisenberg-Weyl group.  This means that one can construct the homogeneous space
$\HW{4}/\Group{U(1)}=: HW(4)$ to remove the redundant group parameter $\kappa$.
According to the Stone-von Neumann theorem, any two unitary irreducible
representations of the $\HW{4}$ group are equivalent, and the group parameter
$\kappa$ leads to the same states, in its action in $\StateSpace{K}$. The
elements of the space $HW(4)$ are represented by the following unitary operators
\begin{equation}
\label{gOperators}
g(p,x)
=\exp\left(\frac{i}{\hbar}(p_\mu \hat{Q}^\mu-x^\mu \hat{P}_\mu)\right)
\,.
\end{equation}
The multiplication law for operators \eqref{gOperators} reads
\begin{equation}
g(p,x)g(\tilde{p},\tilde{x})=
\exp\left(-\frac{i}{2\hbar}(x^\mu \tilde{p}_\mu-p_\mu
\tilde{x}^\mu)\hat{I}\right)  g(p+\tilde{p},x+\tilde{x}) \, .
\end{equation}
The unit operator can be identified with
\begin{equation}
\label{gIdentity}
\UnitOp=g(0,0) \,.
\end{equation}
and the inverse operator reads
\begin{equation}
\label{gInverse}
g^{-1}(p, x)=g(-p,-x) \, .
\end{equation}
The unitary irreducible representation of the group $\HW{4}$ on the
Hilbert space $\StateSpace{K}$ is determined by the following action
\begin{equation}
\label{HW4UnitRep1}
\hat{\mathcal{U}}(\kappa;\,p,x)\psi(\xi)
=\exp(i\kappa)\,\hat{\mathcal{U}}(p,x)\psi(\xi) \,,
\end{equation}
where
\begin{equation}
\label{rep1}
\hat{\mathcal{U}}(p,x)\psi(\xi)=\exp\left(\frac{-ip_\mu x^\mu}{2\hbar}\right)
\exp\left(\frac{ip_\mu \xi^\mu}{\hbar}\right)\psi(\xi-x) \,.
\end{equation}
%


\subsection{Coherent states}

The coherent states, $|p,x\rangle \in \mathcal{K} := L^2(\RNumb^4,\;d^{4}\xi)$,
are constructed as follows
\begin{equation}
\label{rep2}
\Ket{p,x}=\hat{\mathcal{U}}(p,x)\Ket{\Phi_0},~~~\BraKet{\xi}{p,x}
=\hat{\mathcal{U}}(p,x) \BraKet{\xi}{\Phi_0}
=\hat{\mathcal{U}}(p,x)\Phi_0(\xi)\, .
\end{equation}
where $\Phi_0(\xi) \colon \RNumb^4 \rightarrow \mathbb{C}$ is the so-called
fiducial vector and $|\Phi_0 \rangle \in \mathcal{K}$ such that
$\BraKet{\Phi_0}{\Phi_0} = 1$. The freedom in the choice of the fiducial vector
is a powerful feature of the IQ. In fact, the fiducial vector can be treated as
a ``parameter'' of that quantization method.

In what follows we use the notation
\begin{equation}
\label{NotePhi0}
\Ket{\Phi_0}=\hat{\mathcal{U}}(0,0)\Ket{\Phi_0}=\Ket{0,0} \,.
\end{equation}
Due to equation
\begin{eqnarray}
&&\hat{\mathcal{U}}^{-1}(p,x)\hat{\mathcal{U}}(\tilde{p},
\tilde{x})\psi(\xi)=
\hat{\mathcal{U}}(-p,-x)\hat{\mathcal{U}}(\tilde{p},\tilde{x})\psi(\xi) \nonumber
\\
&&=\exp\left(-\frac{i}{2\hbar}(-x^\mu\tilde{p}_\mu+p_\mu\tilde{x}^\mu)\right)
\hat{\mathcal{U}}(\tilde{p}-p,\tilde{x}-x)\psi(\xi) \, , \label{mult}
\end{eqnarray}
we have
\begin{eqnarray}
&&\BraKet{p, x}{\tilde{p}, \tilde{x}}=
\Bra{\Phi_0}\hat{\mathcal{U}}^{-1}(p,x)\hat{\mathcal{U}}(\tilde{p},
\tilde{x})\Ket{\Phi_0} \nonumber \\
&&\nonumber =\exp\left(-\frac{i}{2\hbar}(-x^\mu\tilde{p}_\mu
+p_\mu\tilde{x}^\mu)\right)
\Bra{\Phi_0}\hat{\mathcal{U}}(\tilde{p}-p,\tilde{x}-x)\Ket{\Phi_0} \\
&&=\exp\left(-\frac{i}{2\hbar}(-x^\mu\tilde{p}_\mu+p_\mu\tilde{x}^\mu)\right)
\BraKet{0,0}{\tilde{p}-p, \tilde{x}-x} \, .
\end{eqnarray}
Making the group structure in the above formula explicit, we get
\begin{eqnarray}
& \BraKet{p, x}{\tilde{p}, \tilde{x}} &
=\Bra{\Phi_0}
\hat{\mathcal{U}}\left((p,x)^{-1}\circ(\tilde{p},
\tilde{x})\right)
\Ket{\Phi_0}=\BraKet{0,0}{(p,x)^{-1}\circ(\tilde{p},
\tilde{x})} \nonumber \\
&&
=\Bra{\Phi_0}
\hat{\mathcal{U}}^{-1}\left((\tilde{p},
\tilde{x})^{-1}\circ(p,x)\right)
\Ket{\Phi_0}=\BraKet{(\tilde{p},
\tilde{x})^{-1}\circ(p,x)}{0,0} \, . \label{eq:BraKet2}
\end{eqnarray}
Using \eqref{eq:BraKet2} we  obtain the following invariance property
\begin{eqnarray}
&&\BraKet{(p',x')\circ(p, x)}{(p',x')\circ(\tilde{p}, \tilde{x})}=
\BraKet{p, x}{\tilde{p}, \tilde{x}} \label{BraKet3} \, .
\end{eqnarray}
Since the representation is irreducible, the operators
$\Ket{p,x}\Bra{p,x} : \StateSpace{K} \rightarrow \StateSpace{K}$ satisfy
\begin{equation}\label{resolution}
\frac{1}{A_{\Phi_0}}\int_{\RNumb^{8}}\, d\rho(p,x) \, \Ket{p,x}\Bra{p,x}
=\UnitOp \, ,
\end{equation}
where $A_{\Phi_0}$ is the normalization coefficient.

The Heisenberg-Weyl quantization consists in ascribing uniquely to each point of
the phase space $T^\star \mathcal{M}$ the projection operator
\begin{equation}\label{acs5}
  \RNumb^{8} \ni (p,x) \longrightarrow \Ket{p,x}\Bra{p,x}  \, .
\end{equation}
In the IQ Eq.~\eqref{resolution} is used for mapping
(quantization) of almost any classical observable
$f \colon \RNumb^{8} \rightarrow \RNumb$ onto an operator
$\hat{f} \colon \StateSpace{K} \rightarrow \StateSpace{K}$ as follows
\begin{equation}
\label{mapping}
f \longrightarrow \hat{f} :=
\frac{1}{A_{\Phi_0}}\int_{\RNumb^{8}}\, d\rho(p,x)
\Ket{p,x}f(p,x)\Bra{p,x} \, .
\end{equation}
Inserting the formulas
\begin{equation}\label{sim1}
  \int_{\RNumb^4}d^4\xi\,\, |\xi\rangle\langle \xi| = \UnitOp,~~~~~
  \langle \xi'' | \xi'\rangle = \delta^4 (\xi'' - \xi') \, ,
\end{equation}
where $\delta^4(\xi'' - \xi') = \delta(\xi''^0 - \xi'^0) \delta(\xi''^1 - \xi'^1) \delta(\xi''^2 - \xi'^2) \delta(\xi''^3 - \xi'^3)$, into \eqref{resolution}, and using \eqref{rep1} and \eqref{rep2}, as well as the expression
\begin{equation}\label{sim2}
  \int_{\RNumb} dp \, \exp \left( \frac{ipx}{\hbar} \right) = 2\pi\hbar\, \delta (x) \, ,
\end{equation}
one can easily show that
\begin{equation}\label{sim3}
 A_{\Phi_0} = (2\pi\hbar)^4 \,.
\end{equation}
Thus, this coefficient does not depend on the choice of the fiducial
vector $|\Phi_0\rangle \in \mathcal{K}$, which is not the case while applying
the IQ based on other groups (see, e.g.,
\cite{ast2}--\cite{met} for more details).

Mapping \eqref{mapping} leads to a symmetric operator which, in general, is not
self-adjoint. In fact, a symmetric operator can have many self-adjoint
extensions or none at all \cite{Reed}. This feature makes the integral
quantization non unique, which is undesirable. To solve this problem, we propose
to use POVM type operators which make that mapping unique.

\section{Positive operator valued measures and applications}
\label{POVM}

First, let us define a general form of POVM operators and the corresponding sesquilinear forms.
Let $\mathcal{L}(\StateSpace{K})$ represent a set of bounded linear operators on
the Hilbert space $\StateSpace{K}$. Let $\Omega \subset \RNumb$ denote a set of
allowed values of a quantum observable $A$ and let $\mathcal{F}$ be the
$\sigma$-algebra of subsets of $\Omega$, with $(\Omega,\mathcal{F})$ being a
measurable space. The mapping
$E \colon \mathcal{F} \to \mathcal{L}(\StateSpace{K})$ is called the positive
operator valued measure, POVM, if $E$ fulfils the following conditions:
\begin{itemize}
\item $\forall_{X \in \mathcal{F}}$, $E(X)$ is positive semi-definite;
\item if $\{X_k\}$ is a countable collection of disjoint sets, then
\begin{equation*}
E\left( \bigcup\limits_{k}\, X_{k} \right) =\sum_k \, E(X_k) \, ,
\end{equation*}
and this series converges in the weak topology;
  \item  $E(\emptyset)=0$ and $E(\Omega)=\UnitOp$.
\end{itemize}
The first condition ensures positivity of the quantum probability
\begin{equation}
\label{ProbPOVM}
\Prob(E(X);\hat{\gamma})=\Trace(E(X)\hat{\gamma}) \, ,
\end{equation}
where $\hat{\gamma}$ denotes a density operator, i.e., the state of the quantum
system under consideration. $\Prob(E(X);\hat{\gamma})$ is interpreted as the
probability that values of the observable $E$ belong to the set $X$ (see the so
called ``minimal interpretation of quantum mechanics'' \cite{Busch1996}). The
second condition represents additivity of the measure for disjoint sets and the
probability for mutually exclusive events. The last condition normalizes the
probability to unity.

There is a well-known one-to-one correspondence between bounded sesquilinear forms and bounded linear operators
defined on a Hilbert space \cite{Conway1990}, which may be understood as a precise formulation of the statement that any
linear operator acting in a Hilbert space can be prescribed by its matrix elements.
In fact, the assumption that the forms and operators should be bounded can be partially relaxed---see, for instance,
Theorem 3.5.1 in \cite{Mlak1972}. In either case, sesquilinear forms can be used to study linear operators,
but they can also be used to define quantum observables.

A sesquilinear form in a Hilbert space $\StateSpace{K}$ is defined as a map $ \check{h} \colon \StateSpace{K}
\times \StateSpace{K} \to \CNumb$, fulfilling the following conditions:
\begin{eqnarray}
\label{SesqForm}
&&\check{h}(\psi_2+\psi_2',\psi_1+\psi_1')
= \check{h}(\psi_2,\psi_1)+\check{h}(\psi_2,\psi_1')
+\check{h}(\psi_2',\psi_1)+\check{h}(\psi_2',\psi_1') \, , \nonumber \\
&& \check{h}(\alpha \psi_2,\beta \psi_1)
= \alpha^\star \beta \check{h}(\psi_2,\psi_1) \, ,
\end{eqnarray}
for all $\psi_1, \psi_1',\psi_2, \psi_2' \in \StateSpace{K}$ and all $\alpha, \beta \in \CNumb$.
To construct the POVM operator corresponding to a classical observable $f(p,x)$,
we define the following sesquilinear form
\begin{equation}
\label{POVMSesq}
\check{M}_f(U;\psi_2,\psi_1)
:= \frac{1}{(2\pi\hbar)^4} \int_{\RNumb^{8}} d\rho(p,x) \,
\BraKet{\psi_2}{p,x} \chi(f(p,x) \in U) \BraKet{p,x}{\psi_1} \, ,
\end{equation}
where $\chi(S)=1$, iff relation $S$ is satisfied and $\chi(S) = 0$, otherwise. The
condition $\chi(f(p,x) \in U)=1$ restricts the part of the phase space
transformed by the classical observable $f$ to the set $U \subset \RNumb$.
Using the bra-ket notation, the corresponding operator can be written as
\begin{equation}
\label{POVMOp}
\hat{M}_f(U)
= \frac{1}{(2\pi\hbar)^4} \int_{\RNumb^{8}} d\rho(p,x) \,
\Ket{p,x} \chi(f(p,x) \in U) \Bra{p,x} \,.
\end{equation}
This operator is bounded and self-adjoint (see Appendix~\ref{appA}). Operators of this type
fulfil all conditions defining POVM operators, where $U$ belongs to the
$\sigma$-algebra $\mathcal{F}$.

We say that the full set of operators \eqref{POVMOp} represents the quantum
observable $\hat{f}$ corresponding to a classical observable $f$.  All
required physical characteristics of the observable $\hat{f}$ can be obtained by means
of operators constructed as functions of measures \eqref{POVMOp}. This includes
the expectation value, the variance, and matrix elements of $\hat{f}$.

In the traditional approach, an observable is described by a single operator. A synthesis
of operator measures and values of a given observable is given on the basis of
the spectral theorem \cite{Riesz1956,Byron1970}. Using this idea, in our case
for non-orthogonal measures, one can write the approximate representation of the operator
$\hat{f}$ as
\begin{equation}
\label{ApproxOperf}
\hat{f}(\epsilon;a,b):=\sum_{k} \bar{u}_{k} \, \hat{M}_f(Q_k) \,,
\end{equation}
where $(a < \dots < u_k < u_{k+1} < u_{k+2}< \dots \leq b)$ is a partition of
the interval $ (a,b] \subset\RNumb$, $Q_k=(u_{k},u_{k+1}]$ and
$\bar{u}_k \in Q_k$. $a < b$ determine the range and
$\epsilon = \max_k|u_{k+1}-u_{k}|$ describes the resolution of this observable. In
general, one needs to calculate the weak operator limit of self-adjoint
operators $\hat{f}(\epsilon;a,b)$ for $\epsilon \to 0$ and then take the limit $a \to -\infty$
and $b \to +\infty$. If this limit exists, it is represented by the integral
\begin{equation}
\label{SpectralRepf}
\Bra{\psi_2}\hat{f}\Ket{\psi_1}
=\lim \Bra{\psi_2}\hat{f}(\epsilon;a,b)\Ket{\psi_1}
= \int_{\RNumb} u  d\Bra{\psi_2} \hat{M}_{f}(u) \Ket{\psi_1} \,,
\end{equation}
where
\begin{equation}
\label{DystrybuantaOpMf}
\hat{M}_f(u):= \frac{1}{(2\pi\hbar)^4} \int_{\RNumb^{8}} d\rho(p,x) \,
\Ket{p,x} \chi(f(p,x) \leq u)\Bra{p,x} \,.
\end{equation}
%

A link between measures and quantum physics is established by the fundamental formula
\begin{equation}
\label{ProbPOVMOp}
\Prob(M_f(U);\hat{\gamma})
= \frac{1}{(2\pi\hbar)^4} \int_{\RNumb^{8}} d\rho(p,x) \,
\chi(f(p,x) \in U) \Trace(\Ket{p,x}\Bra{p,x}\, \hat{\gamma})  \, ,
\end{equation}
which describes the probability that the observable $\hat{f}$ has its value in the
set $U$ and the system is described by the density operator $\hat{\gamma}$.

Instead of dealing with symmetric sesquilinear forms, it is often more convenient
to work with corresponding quadratic forms or, in the quantum context, the expectation values.
We use the notation $\Aver{\hat{f};\psi} := \check{f}(\psi,\psi) = \Bra{\psi} \hat{f} \Ket{\psi}$.
The original sesquilinear form can be always recovered by the standard polarization identity \cite{Riesz1956, Mlak1972}
\begin{equation}
\label{Polarization}
\Bra{\psi_2}\hat{f}\Ket{\psi_1}=\frac{1}{4}
\left( \Aver{\hat{f};\psi_1+\psi_2}-\Aver{\hat{f};\psi_1-\psi_2}
+i\Aver{\hat{f};\psi_1+i\psi_2} - i\Aver{\hat{f};\psi_1-i\psi_2} \right) \,.
\end{equation}

In the following we consider only pure states $\Ket{\psi}$, i.e.,
$\hat{\gamma}=\Ket{\psi}\Bra{\psi}$. A  generalization to mixed states is
straighforward. Following definition \eqref{ApproxOperf}, the expectation value of the observable
$\hat{f}$ can be written as a limit of the following sum:
\begin{eqnarray}
\label{ExpectValMf}
&& \Aver{\hat{f};\psi}:=  \lim \Aver{\hat{f}(\epsilon;a,b);\psi}
= \lim \sum_{k} \bar{u}_{k} \, \Aver{\hat{M}_f(Q_k);\psi} \nonumber \\
&& =  \lim \sum_k \bar{u}_k\,
[\Prob(\hat{M}_f(u_{k+1});\psi) -\Prob(\hat{M}_f(u_{k});\psi)] \nonumber \\
&& =\int_{\RNumb} u\, d\Bra{\psi} \hat{M}_{f}(u) \Ket{\psi}\,,
\end{eqnarray}
where $Q_k=(u_{k+1},u_{k}]$ and $\bar{u}_k \in Q_k$. This limit means that the
length of the largest subinterval $Q_k=(u_k,u_{k+1}]$ in the sum
\eqref{ExpectValMf} approaches zero, for every partition
$(a < \dots < u_k < u_{k+1} < u_{k+2}<\dots \leq b)$ of the interval
$(a,b] \subset \RNumb$. Subsequently, one needs to take limits $a \to -\infty$ and
$b \to +\infty$.

A differentiation of the expectation value of operator \eqref{DystrybuantaOpMf}
with respect to $u$ gives the probability density that the observable $\hat{f}$ has the
value $u$:
\begin{equation}
\label{DensProbOpMf}
\frac{\partial}{\partial u} \Aver{\hat{M}_f(u);\psi}
= \frac{1}{(2\pi\hbar)^4} \int_{\RNumb^{8}} d\rho(p,x)
\, \delta(u-f(p,x))|\BraKet{p,x}{\psi}|^2 \,,
\end{equation}
where the Dirac delta distribution is used as the derivative of the step function.
Next, the condition $\bar{u}_k \in Q_k$ in \eqref{ExpectValMf} implies that for every
$k$ there is a point $(p(k),x(k))$ for which $f(p(k),x(k))=\bar{u}_k$, and
formula \eqref{ExpectValMf} can be rewritten as
\begin{equation}
\label{ExpectValMf01}
\Aver{\hat{f};\psi}:= \lim \sum_k f(p(k),x(k)) \, |\BraKet{p(k),x(k)}{\psi}|^2
\,.
\end{equation}
If the above limit exists, the corresponding sesquilinear form can be written as
%
\begin{equation}
\label{ExpectValMf02}
\check{f}(\psi_2,\psi_1)= \frac{1}{(2\pi\hbar)^4} \int_{\RNumb^{8}} d\rho(p,x) \,
\BraKet{\psi_2}{p,x} f(p,x) \BraKet{p,x}{\psi_1} \,.
\end{equation}

Theorem 3.5.1 of \cite{Mlak1972} allows one to associate with the sesquilinear form (\ref{ExpectValMf02}) a unique operator
\begin{equation}
\label{ExpectValMf03}
\hat{f}= \frac{1}{(2\pi\hbar)^4} \int_{\RNumb^{8}} d\rho(p,x) \,
\Ket{p,x} f(p,x) \Bra{p,x} \, ,
\end{equation}
defined on the domain $\mathcal{D} \subset \StateSpace{K}$ consisting of all
$\Ket{\psi_1}$ for which there exists a state dependent finite constant
$m(\psi_1) \geq 0$ such that for all $\Ket{\psi_2}$ the following inequality is
satisfied:
\begin{equation}
\label{SFormOper}
|\check{f}(\psi_2,\psi_1)| \leq m(\psi_1)\Norm{\psi_2} \,.
\end{equation}
This operator does not have to be bounded or self-adjoint, but it is symmetric in $\mathcal{D}$, defined by \eqref{SFormOper}. It can be shown that condition \eqref{SFormOper} is fulfilled if
$f(p,x)\psi_1(p,x)$ is a square integrable function.

Finally, note that since symmetric sesquilinear forms and expectation values are related by the polarization identity (\ref{Polarization}), a sesquilinar form associated with an expectation value (a quadratic form) is well defined, provided that the latter is also well defined (see Appendix \ref{appB}).

The above reasoning shows a limitation of the standard integral quantization
formula \eqref{mapping}, which generally leads to a symmetric operator, which in turn may not always allow for a unique extension to a self-adjoint operator.

Matrix elements of any quantum observable correspond to special values of the associated
sequilinear form. Every function of the quantum observable can be written in
terms of these matrix elements. If a calculation based on operators
\eqref{mapping} fails, one can resort to POVM operators. Their matrix
elements are well determined. In the following we use POVM operators to calculate
appropriate matrix elements, even if we do not show this explicitly.

\subsection{Elementary observables}

The elementary observables are operators corresponding to group elements
representing points of the configuration space. In practice one needs to
construct operators corresponding to a given parametrization of the group
$\HW{4}$.

In the canonical approach to quantization, the generators of the group
$\HW{4}$, i.e., $\hat{Q}^\mu$ and $\hat{P}_\mu$ are considered to be momentum
and position operators. In our approach to the $\HW{4}$ integral quantization
the corresponding operators are defined by the appropriate measures
\eqref{DystrybuantaOpMf} as follows
\begin{eqnarray}
&&\hat{M}_{p_\mu}(u):=
\frac{1}{(2\pi\hbar)^4} \int_{\RNumb^{8}} d\rho(p,x) \,
\Ket{p,x}\chi(p_\mu \leq u) \Bra{p,x} \, , \label{POVMp} \\
&&\hat{M}_{x^\mu}(v):=
\frac{1}{(2\pi\hbar)^4} \int_{\RNumb^{8}} d\rho(p,x) \,
\Ket{p,x}\chi(x^\mu \leq v) \Bra{p,x} \label{POVMx} \, .
\end{eqnarray}
According to \eqref{SpectralRepf}, the operators $\hat{p}_\mu$ and $\hat{x}^\mu$
are defined by the following matrix elements in the state space
\begin{eqnarray}
&&\Bra{e_n} \hat{p}_\mu \Ket{e_m} :=
\int_{\RNumb} u \, d\,\Bra{e_n}\hat{M}_{p_\mu}\Ket{e_m} \, , \label{pHW4} \\
&&\Bra{e_n}\hat{x}^\mu \Ket{e_m} :=
\int_{\RNumb} u \, d\,\Bra{e_n}\hat{M}_{x^\mu}\Ket{e_m} \label{xHW4} \,,
\end{eqnarray}
where vectors $\Ket{e_k}$ form an arbitrary orthonormal basis in the state
space.

It is interesting to compare generators of the group $\HW{4}$ and
quantized operators of momenta and positions.
For this purpose, it is easiest to compare corresponding matrix elements
in a well chosen basis of the carrier space $\StateSpace{K}$.

In the case of momenta operators we choose the eigenbases of the generators
$\hat{P}_\mu$,
\begin{equation}
\label{MatElemP}
\Bra{\eta_{p''}} \hat{P}_\mu \Ket{\eta_{p'}}= \delta^4(p''-p')p'_\mu \, .
\end{equation}
 It is easy to show that
\begin{equation}
\label{Ham3}
\BraKet{\xi}{\eta_p} =: \eta_p(\xi)
= \left( \frac{1}{\sqrt{2\pi \hbar}} \right)^4
\exp \left( i \frac{p_\mu \xi^\mu}{\hbar} \right)
\, .
\end{equation}
In the case of position operators we can work in the eigenbases of the
generators $\hat{Q}^\mu$,
\begin{equation}
\label{MatElemQ}
\Bra{\xi''} \hat{Q}^\mu \Ket{\xi'}= \delta^4(\xi''- \xi')\xi'^\mu \,.
\end{equation}
Straightforward calculations allow to obtain required matrix elements of the
corresponding measures
\begin{eqnarray}
\Bra{\eta_{p''}} \hat{M}_{p_\mu}(u) \Ket{\eta_{p'}} & = &
\delta^4(p''-p') \int_{\RNumb^{4}} d^4p   \,
\chi(p_\mu \leq u) |\tilde{\Phi}_0(p'-p)|^2 \label{ElMacPOVMp} \, ,\\
\Bra{\xi''} \hat{M}_{x^\mu}(u) \Ket{\xi'} & = &
\delta^4(\xi''-\xi') \int_{\RNumb^{4}} d^4x   \,
\chi(x^\mu \leq u) |\Phi_0(\xi'-x)|^2 \label{ElMacPOVMx} \,,
\end{eqnarray}
where the Fourier transform of the fiducial vector is defined as
\begin{equation} 
\label{Ham8}
\tilde{\Phi}_0 (p) := \BraKet{\eta_p}{\Phi_0}
= \frac{1}{(2\pi\hbar)^2} \int_{\RNumb^{4}}d^4 \xi\,
\exp \left( -i \frac{p_\mu\xi^\mu}{\hbar} \right) \Phi_0 (\xi) \,.
\end{equation}
To show (\ref{ElMacPOVMp}) and (\ref{ElMacPOVMx}) we use the expressions (see
\eqref{rep2} and \eqref{rep1})
\begin{equation}\label{Ham6}
\BraKet{\xi}{ p,x}
= \exp \left( - i \frac{p_\mu x^\mu}{2 \hbar} \right)\,
\exp \left( i \frac{p_\mu \xi^\mu}{ \hbar} \right) \Phi_0 (\xi - x)\, ,
\end{equation}
and
\begin{equation}\label{Ham7}
\BraKet{\eta_k}{p,x} = \int_{\RNumb^{4}}d^4\,\xi\,
\BraKet{\eta_k}{\xi} \BraKet{\xi}{ p,x} =
\exp \left( i\, \frac{(p_\mu - 2 k_\mu) x^\mu}{2\hbar}  \right) \,
\tilde{\Phi}_0(k - p) \, ,
\end{equation}
where in the last formula we use Eq.\ (\ref{Ham3}). Finally, repeating the
general calculation of Sec.\ IV, one obtains
\begin{eqnarray}
\Bra{\eta_{p''}} \hat{p}_\mu \Ket{\eta_{p'}} & = &
\delta^4(p''-p') \left(p'_\mu  - \int_{\RNumb^4} d^4p\, p_\mu
|\tilde{\Phi}_0(p)|^2  \right)\label{ElMacpOper}\, ,\\
\Bra{\xi''} \hat{x}^\mu \Ket{\xi'} & = &
\delta^4(\xi''-\xi')\left(\xi'^\mu - \int_{\RNumb^{4}} d^4\xi\,
\xi^\mu |\Phi_0(\xi)|^2 \right) \label{ElMacxOper}\, .
\end{eqnarray}
In accordance with the condition (\ref{SFormOper}) and the subsequent commentary,
the above operators are uniquely defined in the bases
$\Ket{\eta_p}$ and $\Ket{\xi}$ if functions $p_\mu\tilde{\Phi}_0(p)$ and   $\xi^\mu\Phi_0(\xi)$, respectively,  are square
integrable functions.
We see that the generators of the group $\HW{4}$ and corresponding quantized
momenta and positions coincide up to some constants. The constants depend
only on the fiducial vector. Assuming additionally that the fiducial vector has  good parity,
i.e.,
\begin{equation}
\label{Phi0Parity}
\Phi_0(-\xi) = \pm \Phi_0(\xi) \, ,
\end{equation}
one can easily show that these constants are equal to zero.  This means that the
$\HW{4}$ integral quantization method reproduces the canonical momentum and
position operators.


\subsection{Expectation values of the momentum and position operators}

In our approach the coherent states are viewed as points of the quantum
configuration space. This suggests a compatibility of their parametrization
with expectation values of $\hat{p}_\mu$ and $\hat{x}^\mu$ operators, namely
\begin{eqnarray}
\label{CompatCond}
&&\Bra{\tilde{p},\tilde{x}}\hat{p}_\nu\Ket{\tilde{p},\tilde{x}}
=\tilde{p}_\nu  \, , \\
&&\Bra{\tilde{p},\tilde{x}}\hat{x}^\nu\Ket{\tilde{p},\tilde{x}}
=\tilde{x}^\nu \,.
\end{eqnarray}
To fulfil the above compatibility conditions, one needs to choose an appropriate
group parametrization and an appropriate fiducial vector.  It turns out that in
our case the good parity fiducial vector \eqref{Phi0Parity} allows to
satisfy properties \eqref{CompatCond}. This is due to the fact that
\begin{equation}
\label{pMatElemCohState}
\Bra{\tilde{p},\tilde{x}} \hat{p}_\nu \Ket{\tilde{p},\tilde{x}}
 =\tilde{p}_\nu
+ \frac{1}{(2\pi\hbar)^4}\int_{\RNumb^{8}}\, d\rho(p,x)\, p_{\nu}
\left|\int_{\RNumb^{4}} d^{4}\xi\, \Phi_0^\star(\xi)
\exp\left(\frac{ip_\mu\xi^\mu}{\hbar}\right) \Phi_0(\xi-x)\right|^2
\end{equation}
and
\begin{equation}
\Bra{\tilde{p},\tilde{x}} \hat{x}^\nu \Ket{\tilde{p},\tilde{x}} \\
=\tilde{x}^\nu
+ \frac{1}{(2\pi\hbar)^4}\int_{\RNumb^{8}}\, d\rho(p,x) x^{\nu}
\left|\int_{\RNumb^{4}} d^4\xi \Phi_0^\star(\xi)
\exp\left(\frac{ip_\mu\xi^\mu}{\hbar}\right)\Phi_0(\xi-x)\right|^2 \,.
\end{equation}
Due to \eqref{Phi0Parity} the  r.h.s integrals are equal to zero.

To complete our derivation, one can check that the $\HW{4}$ generators have the
same expectation values for the fiducial vectors
satisfying \eqref{Phi0Parity}
\begin{equation}
\Bra{\tilde{p},\tilde{x}}\hat{P}_\nu\Ket{\tilde{p},\tilde{x}}
=\tilde{p}_\nu-i\hbar\int_{\RNumb^4}\, d^{4}\xi \, \Phi^\star_0(\xi)
\frac{\partial}{\partial \xi^\nu}\Phi_0(\xi)
\end{equation}
and
\begin{equation}
\Bra{\tilde{p},\tilde{x}}\hat{Q}^\nu\Ket{\tilde{p},\tilde{x}}
= \tilde{x}^\nu+\int_{\RNumb^4}\, d^{4}\xi \, \xi^\nu \, |\Phi_0(\xi)|^2 \,.
\end{equation}
Again, the parity condition \eqref{Phi0Parity} makes the r.h.s. integrals
vanishing.

We emphasise that the operators $\hat x^\mu$ and $\hat p_\nu$ (as well as $\hat Q^\mu$ and $\hat P_\nu$)
should be understood as characterizing the configuration space rather than referring to a constrained physical
system, such as a particle satisfying a mass-shell condition corresponding to a fixed rest mass $m$.
Such constraints can be imposed by defining suitable transition amplitudes, and we do this in Sec.\ \ref{sec:amplitudes}.
Spectra of each of the operators $\hat x^\mu$ and $\hat p_\nu$  coincide with $\RNumb$. This implies, in particular,
that the operators $\hat{t}:= \hat{x}^0$ and $\hat{p}_0$ are not subject to the well-known argument by Pauli
(see \cite{Eric} and references therein).

\subsection{One dimensional harmonic oscillator }

To compare the IQ based on the Heisenberg-Weyl group with the canonical
quantization, we consider the quantization of a nonrelativistic one-dimensional
harmonic oscillator.
This problem has partially been considered in various ways and contexts (see, for instance,
\cite{Klauder,har1,har2,har3} and references therein).
In the following, we use the POVM formalism to regularize operators obtained
by the quantization procedure. In the present case, the group $\HW{4}$ has to be replaced by
$\HW{1}$. On the other hand, all formulas obtained in previous sections can easily be rewritten and applied.

The classical Hamiltonian of harmonic oscillations reads
\begin{equation}
\label{HOClassHam}
H(p,x)=\frac{p^2}{2m} +\frac{1}{2}m \omega^2 x^2 \, ,
\end{equation}
where $m$ represents the mass and $\omega$ is the frequency of the harmonic
oscillator.

To quantize this Hamiltonian with the Heisenberg-Weyl group $\HW{1}$ we need
two operators
\begin{eqnarray}
\label{Operp2s2}
&& \widehat{p^2}:= \int_{\RNumb} u \, d \hat{M}_{p^2}(u) \, , \\
&& \widehat{x^2}:= \int_{\RNumb} v \, d \hat{M}_{x^2}(v) \, ,
\end{eqnarray}
where the corresponding POVM operators are
\begin{eqnarray}
&& \hat{M}_{p^2}(u) = \frac{1}{2\pi\hbar} \int_{\RNumb^2} dp\, dx \,
\Ket{p,x} \chi(p^2\leq u) \Bra{p,x} \label{SpectralOpp2} \,, \\
&& \hat{M}_{x^2}(u) = \frac{1}{2\pi\hbar} \int_{\RNumb^2} dp\, dx \,
\Ket{p,x} \chi(x^2\leq u) \Bra{p,x} \, . \label{SpectralOps2}
\end{eqnarray}
In the following, we assume that the fiducial vector has good parity,
i.e., $\Phi_0(-\xi)=\pm\Phi(\xi)$.
The matrix elements of operator \eqref{SpectralOpp2} within the ``momentum''
basis
\begin{equation}
\BraKet{\xi}{\eta_p} = \eta_{p}(\xi):=\frac{1}{\sqrt{2\pi \hbar}}
\exp \left( \frac{i p \xi}{\hbar} \right)
\end{equation}
are
\begin{equation}
\label{MEPOVMp2MomBas}
\Bra{\eta_{p''}} \hat{M}_{p^2}(u) \Ket{\eta_{p'}} =
  \delta(p''-p')\, \int_{\RNumb} dp\,
\chi(p^2 \leq u) |\tilde{\Phi}_0(p' - p)|^2 \, .
\end{equation}
Using these matrix elements and the decomposition of unity \eqref{EtaUnity}, one
gets required matrix elements for $\hat{M}_{p^2}(u)$ in the ``position'' basis
\begin{equation}
\label{MEPOVMp2PosBas}
\Bra{\xi''} \hat{M}_{p^2}(u) \Ket{\xi'} =
\frac{1}{2\pi\hbar} \int_{\RNumb} dp'\, \int_{\RNumb} dp
\exp \left( i\frac{p'(\xi''-\xi')}{\hbar} \right) \chi \left( (p' - p)^2 \leq u \right) | \tilde \Phi_0(p)|^2 \, .
\end{equation}
Similarly,
\begin{equation}
\label{MEPOVMx2PosBas}
\Bra{\xi''} \hat{M}_{x^2}(v) \Ket{\xi'} =
 \delta(\xi''-\xi')
\int_{\RNumb} dx\, \chi(x^2 \leq v) |\Phi_0(\xi'-x)|^2 \, .
\end{equation}
Finally, the required matrix elements of operators \eqref{Operp2s2} are
\begin{eqnarray}
\label{MatElemOperp2s2}
&&\Bra{\xi''} \widehat{p^2} \Ket{\xi'}
= -\hbar^2 \delta^{(2)}(\xi''-\xi')
+ \delta(\xi''-\xi') \int_{\RNumb}\, dp\, p^2 |\tilde \Phi_0(p)|^2 \,, \\
&&\Bra{\xi''} \widehat{x^2} \Ket{\xi'}
= \delta(\xi''-\xi') \int_{\RNumb} dx\, x^2 |\Phi_0(\xi'-x)|^2 \, ,
\end{eqnarray}
where the parity of the fiducial vector and the formula for the second
derivative of the delta distribution
\begin{equation}
\label{SecDerDelta}
\frac{1}{2\pi \hbar} \int_{\RNumb} dp\, p^2
\exp\left(i \frac{p(\xi''-\xi')}{\hbar}   \right)
=-\hbar^2 \delta^{(2)}(\xi''-\xi')
\end{equation}
were used.

Using these matrix element as the integral kernel of the quantized Hamiltonian,
we get
\begin{equation}
\label{HarmOscAction}
\int_{\RNumb} d\xi'\, \Bra{\xi} \hat{H} \Ket{\xi'} \psi(\xi')
=\left\{-\frac{\hbar^2}{2m} \frac{d^2}{d\xi^2} + \frac{1}{2} m \omega^2 \xi^2 \right\}
\psi(\xi) + C \psi(\xi) \, ,
\end{equation}
where $C$ is a constant dependent only on the fiducial vector
\begin{equation}
\label{HarmOscConst}
C= \frac{1}{2m} \int_{\RNumb} dp\, p^2 |\tilde \Phi_0(p)|^2
+\frac{1}{2} m \omega^2 \int_{\RNumb} dx\, x^2 |\Phi_0(x)|^2 \, .
\end{equation}
Expression \eqref{HarmOscAction} implies that the $\HW{1}$ quantization,
that uses the  POVM operators, reproduces the quantum harmonic oscillator
Hamiltonian as obtained within  the canonical quantization scheme:
\begin{equation}
\label{QuantHarmOscHam}
\hat{H}
=-\frac{\hbar^2}{2m} \frac{d^2}{d\xi^2} + \frac{1}{2} m \omega^2 \xi^2 + C \, .
\end{equation}
An open problem is related to the additional constant term $C$ and its physical
meaning.  In principle, every classical Hamiltonian $H+C$ leads to the same
Hamilton equations, independently of value of $C$. This suggests that such a
constant term has no meaning. On the other hand, quantization of other
observables can be sensitive to $C$, as it happens in the case of elementary
observables $p_\mu$ and $x^\mu$.

\section{Relativistic models}

\subsection{Poincar\'e covariance}

In this section we show that our quantization scheme can be applied to
  relativistic models. We start by checking its covariance properties with
  respect to the Poincar\'e group. This requires a few elements, which we now
  shortly describe.

  According to our assumptions, every point of the configuration space is
  labelled by a pair of four-vectors, and it transforms under Poincar\'e
  transformations $w(a,\Lambda)$ as
  $w(a,\Lambda)(p,x)=(p',x'):=(\Lambda p, \Lambda x +a)$, where
\begin{equation}
\label{pqLorentz}
p'^{\mu} = \Lambda\indices{^\mu_\nu} p^\nu
\text{ and }
x'^{\mu} = \Lambda\indices{^\mu_\nu} x^\nu +a^\mu  \,.
\end{equation}
Here $\Lambda$ represents Lorentz transformations and $\{a^\mu\}$ parametrizes
four-translations.

Transformation formulas \eqref{pqLorentz} determine the standard action of the
Poincar\'e group on scalar functions on the configuration space as follows:
\begin{equation}
\label{FunLorentz}
\mathcal{G}(a,\Lambda) f(p,x):= f(w(a,\Lambda)^{-1}(p,x)) \,.
\end{equation}
As a next element, we specify the action of the Poincar\'e group in the Hilbert
space $\StateSpace{K}$ in which the unitary representation of the group $\HW{4}$
is also constructed. Without loss of generality, we assume that the variables
$\xi=(\xi^0,\xi^1,\xi^2,\xi^3)$ also represent components of a four-vector in
the Minkowski space. It implies that the scalar product in $\StateSpace{K}$ is
invariant not only with respect to the Heisenberg-Weyl group but also with
respect to the Poincar\'e group, i.e., the measure $d^4\xi$ is invariant with
respect to both groups
\begin{equation}
\label{ScProdHWLorentz}
\BraKet{\mathcal{G}(a,\Lambda) \psi_2}{\mathcal{G}(a,\Lambda) \psi_1}
=\BraKet{\psi_2}{\psi_1} \,.
\end{equation}
Equation (\ref{ScProdHWLorentz}) suggests that in our case the integral
quantization based on the Heisenberg--Weyl group $\HW{4}$ remains covariant with
respect to the Poincar\'e group. To show that this is indeed the
case, consider a general form of a POVM operator \eqref{POVMSesqOp}
corresponding to a scalar observable $f$. Denoting
$f'(p,x)=f(w(a,\Lambda)^{-1}(p,x))$ we see that \eqref{POVMSesqOp} transforms in
a covariant way with respect to the Poincar\'e group:
\begin{eqnarray}
\label{POVMLorentz}
&& \hat{M}_{f'}(U)=\hat{M}_{\mathcal{G}(a,\Lambda)f}(U)=
\frac{1}{(2\pi\hbar)^4} \int_{\RNumb^{8}} d\rho(p,x) \,
\Ket{p,x} \chi(f(w(a,\Lambda)^{-1}(p,x)) \in U) \Bra{p,x} \nonumber \\
&& =\frac{1}{(2\pi\hbar)^4} \int_{\RNumb^{8}} d\rho(p,x) \,
\Ket{w(a,\Lambda)(p,x)} \chi(f(p,x) \in U) \Bra{w(a,\Lambda)(p,x)} \nonumber \\
&& = \hat{\mathcal{G}}(a,\Lambda) \hat{M}_{f}(U)
\hat{\mathcal{G}}(a,\Lambda)^\dagger \,,
\end{eqnarray}
where $\hat{\mathcal{G}}(a,\Lambda)\Ket{p,x}:=\Ket{w(a,\Lambda)(p,x)}$ defines a
representation of the Poincar\'e group in terms of the coherent
states. Expression \eqref{POVMLorentz} shows that a function representing a
classical scalar observable transformed under elements of the Poincar\'e group
leads to a correctly transformed quantum observable. This property is
independent on the form of the fiducial vector $\Phi_0$.

Note that the representation $\hat{\mathcal{G}}(a,\Lambda)$ differs from the
representation defined by the left shift operation in $\StateSpace{K}$, i.e.,
given by $\mathcal{G}(a,\Lambda)\psi(\xi)=\psi(w(a,\Lambda)^{-1}\xi)$. Both
representations coincide if the fiducial vector satisfies the following
condition: $\Phi_0(\xi-w(a,\Lambda) x)=\Phi_0(w(a,\Lambda)^{-1}\xi-x)$. Because
only constant functions are invariant with respect to all translations, this condition cannot, in general, be fulfilled.
However, for
$a=0$ both representations coincide when $\Phi_0$ is invariant with respect to the Lorentz group.


\subsection{Quantum Hamiltonian of a test particle in the Minkowski spacetime}

In what follows we examine the eigenvalue problem of the quantum Hamiltonian
\eqref{gHam} describing the motion of a test particle with the rest mass $m$ in
the Minkowski spacetime. The geodesic Hamiltonian fulfils, due to \eqref{gHam},
the following equation
\begin{equation}\label{Ham1a}
H(p,x) = \frac{1}{2} g^{\mu\nu} p_\mu p_\nu = -\frac{1}{2} m^2 \, .
\end{equation}
The POVM operators of the quantum Hamiltonian $\hat{H}$ have the form
\begin{equation}
\label{POVMHam}
\hat{M}_{H}(u) = \left(\frac{1}{2\pi\hbar}\right)^4
\int_{\RNumb^8} d\rho(p,x) \,
\Ket{p,x} \chi(H(p,x)\leq u) \Bra{p,x} \,.
\end{equation}
The quantum Hamiltonian itself is determined by the following sesquilinear form
\begin{eqnarray}
\label{Ham2}
&& \Bra{\psi_2}\hat{H}\Ket{\psi_1}
= \left(\frac{1}{2\pi\hbar}\right)^4 \int_{\RNumb^{8}}\, d\rho(p,x)
\BraKet{\psi_2}{p,x} H(p,x)\BraKet{p,x}{\psi_1} \nonumber\\
&& =\frac{1}{2} g^{\mu\nu}
\left(\frac{1}{2\pi\hbar}\right)^4 \int_{\RNumb^8} d\rho(p,x) \,
\BraKet{\psi_2}{p,x} p_{\mu}p_{\nu} \BraKet{p,x}{\psi_1} \,.
\end{eqnarray}
The last equality shows that the generalization to four dimensions of the
harmonic oscillator matrix elements \eqref{MatElemOperp2s2} can be directly used
in actual calculations.

In what follows we show that functions $\eta_p(\xi)$ defined in Eq.\
(\ref{Ham3}) are generalized eigenstates of $\hat{H}$ defined by \eqref{Ham2},
if the coherent states $| p,x \rangle \in \mathcal{K}$ are generated from a
suitably chosen fiducial vector $\Ket{\Phi_0} \in \mathcal{K}$. For this
purpose, we need to calculate matrix elements of the Hamiltonian \eqref{Ham2}
within the states $\Bra{\eta_{k^\prime}}$ and $\Ket{\eta_k}$, i.e.,
\begin{equation}
\label{Ham5}
\Bra{\eta_{k^\prime}} \hat{H} \Ket{\eta_k}
=  \frac{1}{(2\pi\hbar)^4} \int_{\RNumb^{8}}\,d^4 p\, d^4 x\,
\BraKet{\eta_{k^\prime}}{ p,x}\,
\frac{1}{2}\, g^{\mu\nu} p_\mu p_\nu\,\BraKet{p,x}{\eta_k} \, .
\end{equation}
Inserting \eqref{Ham7} into \eqref{Ham5} we obtain
\begin{equation}\label{Ham9}
\Bra{\eta_{k^\prime}} \hat{H} \Ket{\eta_k}
= \frac{1}{2}\delta^4 (k^\prime - k)\,\int_{\RNumb^{4}}d^4 p\,
g^{\alpha\beta} p_\alpha p_\beta\,
\tilde{\Phi}_0^\star (k - p) \tilde{\Phi}_0 (k^\prime - p) \, .
\end{equation}
The key element of the further procedure is to use the orthogonal
decomposition of unity in the carrier space $\mathcal{K}$ in terms of the
generalized states \eqref{Ham3}, which reads
\begin{equation}
\label{EtaUnity}
\int_{\RNumb^4} d^4p\, \Ket{\eta_p}\Bra{\eta_p}=\UnitOp \, .
\end{equation}
The validity of \eqref{EtaUnity} results from the theory of Fourier transforms
in the context of distributions (see, e.g., \cite{Strichartz2003}) and is
commonly used in quantum formalisms.

Using the matrix elements of the Hamiltonian \eqref{Ham5} and the decomposition
of the unity \eqref{EtaUnity}, we obtain the following:
\begin{eqnarray}
\label{HamAct}
&&\hat{H}\Ket{\psi}= \int_{\RNumb^4} d^4k' \Ket{\eta_{k'}}
\Bra{\eta_{k'}} \hat{H}\Ket{\psi}
=\int_{\RNumb^4} d^4k'\, \Ket{\eta_{k'}} \int_{\RNumb^4} d^4 k''
\Bra{\eta_{k'}} \hat{H} \Ket{\eta_{k''}} \BraKet{\eta_{k''}}{\psi}
 \nonumber \\
&& = \int_{\RNumb^4} d^4k' \Ket{\eta_{k'}} \int_{\RNumb^4} d^4 k''
\frac{1}{2}\delta^4 (k' - k'')\,
\int_{\RNumb^{4}}d^4p\,
g^{\alpha\beta} p_\alpha p_\beta\, \tilde{\Phi}_0^\star (k''- p)
\tilde{\Phi}_0 (k'- p) \BraKet{\eta_{k''}}{\psi}
 \nonumber \\
&& =\frac{1}{2} g^{\alpha\beta} \int_{\RNumb^4} d^4k' \Ket{\eta_{k'}}
\int_{\RNumb^{4}}d^4 p\, p_\alpha p_\beta\,
|\tilde{\Phi}_0 (k' - p)|^2 \BraKet{\eta_{k'}}{\psi} \,,
\end{eqnarray}
for any $|\psi\rangle \in \mathcal{K}$.

After the change of variables $p \to p+k'$, we get
\begin{eqnarray}
\label{HamAct2}
&&  2 \hat{H}\Ket{\psi}
= \left(\int_{\RNumb^{4}}d^4 p\, |\tilde{\Phi}_0 (-p)|^2\right)
\int_{\RNumb^4} d^4k' g^{\alpha\beta} k'_{\alpha} k'_{\beta}
\BraKet{\eta_{k'}}{\psi}  \Ket{\eta_{k'}}
\nonumber \\
&& + \left( \int_{\RNumb^{4}}d^4 p\,p_\beta |\tilde{\Phi}_0 (-p) |^2 \right)
\int_{\RNumb^4} d^4k' 2 g^{\alpha\beta} k'_{\alpha}
 \BraKet{\eta_{k'}}{\psi}  \Ket{\eta_{k'}} \nonumber \\
&& + \left( \int_{\RNumb^{4}} d^4 p\, g^{\alpha\beta} p_\alpha p_\beta
|\tilde{\Phi}_0 (-p) |^2 \, \right)
\int_{\RNumb^4} d^4k' \BraKet{\eta_{k'}}{\psi}  \Ket{\eta_{k'}} \, .
\end{eqnarray}
Now, assuming $\Ket{\psi}=\Ket{\eta_k}$, we obtain
\begin{equation}
\label{HamAct3}
 2 \hat{H}\Ket{\eta_k} = \tilde{\lambda}_k \Ket{\eta_k} \, ,
 \end{equation}
where the generalized eigenvalues $\tilde{\lambda}_k$ are
\begin{eqnarray}
\label{HamAct4}
&&  \tilde{\lambda}_k
= \left(\int_{\RNumb^{4}}d^4 p\, |\tilde{\Phi}_0 (-p)|^2\right)
 g^{\alpha\beta} k_{\alpha} k_{\beta}
+ \left( \int_{\RNumb^{4}}d^4 p\,p_\beta |\tilde{\Phi}_0 (-p) |^2 \right)
2 g^{\alpha\beta} k_{\alpha}  \nonumber \\
&& + \int_{\RNumb^{4}} d^4 p\, g^{\alpha\beta} p_\alpha p_\beta
|\tilde{\Phi}_0 (-p) |^2 \,.
\end{eqnarray}
Quantization of the right hand side of \eqref{Ham1a} gives the operator
proportional to the unit operator, $-1/2\, m^2 \UnitOp$. Taking into account
that every function is an eigenfunction of the unit operator, one can identify
the eigenvalues \eqref{HamAct4} with $- m^2$, so that we have
\begin{eqnarray}
\label{HamAct5}
&&  \left(\int_{\RNumb^{4}}d^4 p\, |\tilde{\Phi}_0 (-p)|^2\right)
 g^{\alpha\beta} k_{\alpha} k_{\beta}
+ \left( \int_{\RNumb^{4}}d^4 p\,p_\beta |\tilde{\Phi}_0 (-p) |^2 \right)
2 g^{\alpha\beta} k_{\alpha}  \nonumber \\
&& + \int_{\RNumb^{4}} d^4 p\, g^{\alpha\beta} p_\alpha p_\beta
|\tilde{\Phi}_0 (-p) |^2 \, = - m^2 \, .
\end{eqnarray}
In what follows we try to simplify \eqref{HamAct5}. One can easily show that
\begin{equation}\label{Ham12}
\int_{\RNumb^{4}}d^4 p\,|\tilde{\Phi}_0 (-p) |^2
= \int_{\RNumb^{4}}d^4 \xi\,|\Phi_0 (\xi) |^2  = 1\, ,
\end{equation}
as the fiducial vector $\Phi_0 $ is normalized.

If the fiducial vector is an even or odd function of each of its variables,
it is not difficult to find that
\begin{equation}\label{Ham13}
I_\beta := \int_{\RNumb^{4}}d^4 p\,p_\beta |\tilde{\Phi}_0 (p) |^2
= -  I_\beta, \quad \beta = 0,1,2,3,
\end{equation}
so that  $I_\beta = 0$.

Taking into account these two simplifications, we can write \eqref{HamAct5} in the
form
\begin{equation}\label{Ham14}
- m^2 =  g^{\alpha\beta}k_\alpha k_\beta
+ \int_{\RNumb^{4}}d^4 p\,g^{\alpha\beta}p_\alpha p_\beta
|\tilde{\Phi}_0 (p) |^2  \, .
\end{equation}
Further simplifications are possible, depending on the choice of the fiducial
vector in \eqref{Ham14}. Assuming the fiducial vector as the vacuum state of
the four dimensional harmonic oscillator, see App.~\ref{appC},
\begin{equation}\label{Vac4DHo}
\Phi_0 (\xi) =  \prod_{\mu=0}^3
\left(\frac{\lambda_\mu}{\pi\hbar}\right)^{\frac{1}{4}}
\exp\left(-\frac{\lambda_\mu (\xi^\mu)^2}{2\hbar}\right)\,
\end{equation}
with $\lambda_0,\lambda_1=\lambda_2=\lambda_3 >0$ and $\lambda_0=3\lambda_3$, we
obtain
\begin{equation}\label{finally}
g^{\alpha\beta}k_\alpha k_\beta = - m^2 \,.
\end{equation}

Therefore, a suitable choice of the fiducial vector leads to the result that the
Hamiltonian eigenvalues satisfy the relationship quite similar to the relation
among classical momenta. The quantum Hamiltonian $\hat{H}$ has a continuous
spectrum, consisting of infinitely many eigenvalues $- \frac{1}{2} m^2$, each
being infinitely many fold degenerate.  In what follows we assume that the
fiducial vector $|\Phi_0 \rangle$ is chosen to be defined by \eqref{Vac4DHo}.

The assumed form of the fiducial vector has an
additional advantage.  In the context of the presented quantization method,
coherent states are defined as representing the points of the quantum
phase space of our physical system.  In our model, coherent states are used
to represent both four-momentum and four-position simultaneously. It is
therefore important to find the smearing with which the positions and momenta
are localised. This can be specified by the uncertainty principle, which
states that the uncertainty in the position and momentum of a particle is
related by the following equation \cite{Robertson1929}:
\begin{equation}\label{uncert}
\Var{\hat{p}_\mu;\Ket{p,x}}\Var{\hat{x}^\nu;\Ket{p,x}}
\geq \frac{1}{4}\left|\Bra{p,x}[\hat{p}_\mu,\hat{x}^\nu]\Ket{p,x}\right|^2\, .
\end{equation}
For a real valued fiducial vector ($\Phi_0(\xi)\in\RNumb$), the right hand side
of inequality (\ref{uncert}) is equal to
\begin{equation}
\frac{1}{4}\left|\Bra{p,x}[\hat{p}_\mu,\hat{x}^\nu]\Ket{p,x}\right|^2=
\hbar^2\left|\int_{\RNumb^4} d^4\xi\, \xi^\nu \Phi_0(\xi)
\frac{d}{d\xi^\mu}\Phi_0(\xi)\right|^2 \, .
\end{equation}
In case of the fiducial vector in form (\ref{Vac4DHo}), assuming
$\lambda_0=3\lambda_3$, one gets
\begin{equation}
\frac{1}{4}\left|\Bra{p,x}[\hat{p}_\mu,\hat{x}^\nu]\Ket{p,x}\right|^2=\left\{
\begin{array}{lll}
0& \mbox{for}&\nu\neq \mu, \\
\frac{\hbar^2}{4}&\mbox{for}&\nu=\mu.
\end{array}
\right.
\end{equation}
Therefore, the fiducial vector in the adopted form minimizes the uncertainty
principle, independently of parameters $\lambda_0$ and $\lambda_3$.

In Eq.~\eqref{uncert} the variance, which describes a stochastic deviation from
the expectation value of a quantum observable $\hat{A}$ in the quantum state
$|\Psi\rangle \in \mathcal{K}$, is defined as follows
\begin{equation}
\label{var5}
 \Var{\hat{A};|\Psi\rangle}
:= \langle (\hat{A} - \langle\hat{A};\Psi\rangle)^2;\Psi\rangle
= \langle\hat{A}^2;\Psi\rangle - \langle\hat{A};\Psi\rangle^2 \, ,
\end{equation}
where $\langle\hat{B};\Psi\rangle := \langle \Psi|\hat{B}|\Psi\rangle$.

\subsection{Quantum motion and amplitudes in the configuration space}
\label{sec:amplitudes}

In the traditional approach to quantum mechanics, evolution of a quantum system
is enumerated by the external parametric time. For example, in the local field
theory, propagators give rise to probability amplitudes for the time evolution
of a particle from one state to another. In this article, we work in a
configuration space, in which time is treated on equal footing with other
observables. Such a framework supports relativity in a natural way, but it also
introduces many changes with respect to the standard point of view
\cite{PEv2023,PEvClock2024,third}.

Here, we wish to mention two intriguing
experiments, which strongly suggest the existence of interference in time.  They
are analogous to Young's double-slit experiment, but in the time
domain. The first one \cite{Lindler} shows that electrons interfere on two temporal
slits. The analysis of this experiment can be found in \cite{Hor} under the
assumption that time is an observable, not a parameter.
The same effect is observed in the second experiment \cite{Tirole}, but for
photons. It is tempting to treat this interference as being of fundamental nature, similarly to the
interference on spatial slits.
In principle, the IQ model should allow to explain the results of these
experiments by considering the outgoing state of a particle/photon as a superposition
of its states associated with different times.

We are interested in describing the evolution of a system in terms of probability
amplitudes between points in the configuration space. Amplitudes of this type can be
defined for the initial and final states, and they take into account a set of
intermediate states, which in fact characterize the propagating
object. Although a construction of such amplitudes can be described in quite
general terms, we will discuss it here for the simplest case of a test
Klein-Gordon type spinless particle with a given mass $m$.

To avoid standard problems with Dirac delta-like distributions, instead of
working with an exact mass shell, we introduce a mass layer
$\mathcal{J}_{m,\epsilon}$ of thickness $\epsilon$ defined as
\begin{equation} \label{eq:MassLayer}
\mathcal{J}_{m,\epsilon} := \left\{ p ~ \colon
~ -\sqrt{m^2+\mathbf{p}^2+\epsilon} \leq p_0 \leq -\sqrt{m^2+\mathbf{p}^2},~~
\mathbf{p} \in \RNumb^3 \right\} \,.
\end{equation}
For stable particles physical results are obtained by taking the limit
$\epsilon \rightarrow 0$, or in practice a very small $\epsilon$. Definition
\eqref{eq:MassLayer} is, of course, based on solutions of the equation
$g^{\alpha\beta}p_\alpha p_\beta = - m^2$.

The operator $P_{\mathcal{J}_{m,\epsilon}}$ projecting onto
$\mathcal{J}_{m,\epsilon}$ is constructed from the generalized eigenstates
$\Ket{\eta_p}$ of the test particle Hamiltonian. It reads
\begin{equation}
\label{eq:ProjMassLayer}
P_{\mathcal{J}_{m,\epsilon}} :=
\int_{\RNumb^4} d^4 p\,
\Ket{\eta_p} \chi(p \in \mathcal{J}_{m,\epsilon}) \Bra{\eta_p} \, .
\end{equation}
Note that the states $\Ket{\eta_p}$ (or more precisely the wave functions
(\ref{Ham3})) coincide with elementary solutions of the Klein-Gordon equation in
the Minkowski spacetime.

The transition amplitude of the particle of mass
$m\geq 0$ from the state $\Ket{\Psi_1}$ to the state $\Ket{\Psi_2}$ is given by
the following matrix element of the projection operator
\begin{equation}\label{Tran2}
\mathcal{A}_{m,\epsilon}
:=  \Bra{\Psi_2} P_{\mathcal{J}_{m,\epsilon}} \Ket{\Psi_1} \,.
\end{equation}
Choosing $\Ket{\Psi_1}=\Ket{p',x'}$ and $\Ket{\Psi_2}=\Ket{p'',x''}$, one
obtains the propagator of the test particle in the configuration space.

A more detailed analysis of amplitudes of this type will be given in our next
paper \cite{QMin} for the configuration space generated by the $\HW{4}$ group.


\subsection{Quantization of $N$-particle systems}

In all examples discussed so far we have considered ``single particle'' type
problems. In fact, the IQ method is not based on the notion of particles, but rather on
the notion of degrees of freedom of the system under consideration. In order to promote a ``single particle'' description to the $N$-particle framework, one has to find an appropriate group representing
the configuration space of the classical system. Subsequently, one can follow the footsteps described in previous sections.

To be more specific, let us consider an $N$-body system with the configuration
space identified with the group $\Group{G}$. The first task consists in finding an
irreducible representation of that group and construct the appropriate set of
coherent states $\Ket{g}$. Next, one can either use POVM operators
described in this paper or consider a more symbolic form of the quantization formula
\begin{equation}
\label{NBodyQuant}
f(g) \to \hat{f}:= \int_{\Group{G}} d\mu(g) \Ket{g} f(g) \Bra{g} \, ,
\end{equation}
where $g$ denotes group parameters. The simplest form of the many-body group $\Group{G}$ is the direct product $\Group{G}=\bigotimes_{i=1}^N \Group{G}_i$ of single-particle groups $\Group{G}_i$. Here the group $\Group{G}_i$ corresponds to the $i$-th particle.
Such a choice adheres to the traditional construction of many-body spaces as tensor products
of the single-particle spaces. In this case, the resulting coherent states
$\Ket{g}$ are products of single-particle coherent states $\Ket{i;g_i}$,
\begin{equation}
\label{NBodyQuant2}
\Ket{g}=\bigotimes_{i=1}^N \Ket{i;g_i}
=\Ket{1;g_1}\Ket{2;g_2} \dots \Ket{N;g_N}\, ,
\end{equation}
where $g_i$ refers to the parameters of the group $\Group{G}_i$. Quantization of a classical function $f \colon \bigotimes_{i=1}^N \Group{G}_i \to \RNumb$ describing an $N$-body observable can be performed by applying Eq.\ \eqref{NBodyQuant} as follows:
\begin{eqnarray}
\label{NBodyQuant3}
&& f(g_1,g_2,\dots,g_N) \to \hat{f}:=
\int_{\Group{G}_1} d\mu_1(g_1)\int_{\Group{G}_2} d\mu_2(g_2)
\dots \int_{\Group{G}_N} d\mu_N(g_N) \nonumber\\
&&
\Ket{1;g_1}\Ket{2;g_2} \dots \Ket{N;g_N}
f(g_1,g_2,\dots,g_N)
\Bra{1;g_1}\Bra{2;g_2} \dots \Bra{N;g_N} \,.
\end{eqnarray}
If additional constraints on the motion of the particle are present, they can easily be incorporated into the quantization
formula \eqref{NBodyQuant3} by restricting the integration region.

The integral quantization with the group $\bigotimes_{i=1}^N \Group{G}_i$ leads
to a construction of standard many-body spaces such as the Fock space, their
symmetrized and antisymmetrized forms. It allows for an analysis of many body
effects, symmetries, correlations (like Einstein--Podolsky--Rosen) and interactions. However, one
needs to note that this formalism is more flexible and allows one to incorporate
additional features associated with the appropriate choice of the group
$\Group{G} \supset \bigotimes_{i=1}^N \Group{G}_i$ and the fiducial vector. A detailed analysis of many-body aspects of the integral quantization remains beyond the scope of the present paper.

\section{Summary}\label{Con}

In this paper we use the Heisenberg-Weyl group to construct a space
of coherent states.  This group reproduces standard canonical commutation
relations among positions and momenta. In addition, its natural parametrization
is compatible with the Minkowski spacetime.

The carrier space of that representation is used as the Hilbert space of
the considered quantum system. Since the representation is irreducible, there
exists the decomposition of unity in the carrier space that can be used for
mapping of almost any classical observable onto a  symmetric  operator in that
Hilbert space.

The POVM approach is an extension
of the standard approach with the formula \eqref{mapping}. It allows to overcome
the problem that operators \eqref{mapping} are usually only symmetric and may not have
unique  self-adjoint extensions.  The idea is to construct first matrix elements of
required POVM measure corresponding to a classical observable $f$. As the POVM
operators are bounded and therefore self-adjoint, their matrix elements are
defined on the entire state space $\StateSpace{K}$ of the physical system.
Using these matrix elements, one can construct required physical quantities,
such as expectation values, variances, etc, directly from their definitions.
Applying the POVM ideas, we show that our integral quantization reproduces the
result obtained within the canonical quantization approach in the case of a
commonly known harmonic oscillator.

The POVM approach renders the IQ method applicable to quantization of
sophisticated gravitational systems. In particular, to the quantization of
motion of test particles in curved spacetime \cite{QSch}.

\appendix

\section{Selfadjoint measures } \label{appA}

We show that every operator generated by the form \eqref{POVMSesq} is
bounded. To see this, let us consider the corresponding sesquilinear form
\begin{equation}
\label{POVMSesqA}
\check{M}_f(U;\psi_2,\psi_1)=
\frac{1}{(2\pi\hbar)^4} \int_{\RNumb^{8}} d\rho(p,x) \,
\BraKet{\psi_2}{p,x} \chi(f(p,x) \in U) \BraKet{p,x}{\psi_1} \, ,
\end{equation}
where $\psi_1$ and $\psi_2$ belong to the carrier space $\StateSpace{K}$.
As $\chi(f(p,x) \in U) \leq 1$, this form can be bounded as follows:
\begin{equation}
\label{POVMSesq2}
|\check{M}_f(\psi_2,\psi_1)| \leq
\frac{1}{(2\pi\hbar)^4} \int_{\RNumb^{8}} d\rho(p,x) \,
|\BraKet{\psi_2}{p,x}\BraKet{p,x}{\psi_1}| \,.
\end{equation}
Making use of the H\"older inequality
\begin{eqnarray}
\lefteqn{\int_{\RNumb^{8}} d\rho(p,x) \, |\BraKet{\psi_2}{p,x}\BraKet{p,x}{\psi_1}|
\leq } \nonumber \\
&& \left(\int_{\RNumb^{8}} d\rho(p,x) \,
|\BraKet{\psi_2}{p,x}|^2 \right)^{\frac{1}{2}}
\left(\int_{\RNumb^{8}} d\rho(p,x) \,
|\BraKet{p,x}{\psi_1}|^2 \right)^{\frac{1}{2}} \, ,
\label{Hoelder}
\end{eqnarray}
we get
\begin{equation}
\label{POVMSesq3}
|\check{M}_f(\psi_2,\psi_1)| \leq \frac{1}{A_{\phi}}
\left(\int_{\RNumb^{8}} d\rho(p,x) \,
|\BraKet{\psi_2}{p,x}|^2 \right)^{\frac{1}{2}}
\left(\int_{\RNumb^{8}} d\rho(p,x) \,
|\BraKet{p,x}{\psi_1}|^2 \right)^{\frac{1}{2}} \,.
\end{equation}
It can be shown directly, using the
coherent states resolution of unity, that
\begin{equation}
\label{KtoKG}
\BraKet{\psi_2}{\psi_1}= \frac{1}{(2\pi\hbar)^4} \int_{\RNumb^8} d\rho(p,x)
\BraKet{p,x}{\psi_2}^\star \BraKet{p,x}{\psi_2} < \infty \, .
\end{equation}
This shows that every function $\BraKet{p,x}{\psi_k}$, where
$\psi_k \in \StateSpace{K}$, belongs to the space of square integrable functions
$L^2(\RNumb^8,d\rho(p,x))$ and the right hand side of \eqref{POVMSesq3} is
finite for every $\psi_1,\psi_2 \in \StateSpace{K}$.

The inequality \eqref{POVMSesq3} is a required and sufficient condition that the
operator generated by the sequilinear form (\ref{POVMSesq}) is bounded \cite{Conway1990} (see also
Theorem 3.5.2 of \cite{Mlak1972}).

Since the sequilinear form \eqref{POVMSesq} is symmetric, the generated operator
is also symmetric, and because it is bounded, it is self-adjoint. In the
bra-ket notation this operator can be shortly written as
\begin{equation}
\label{POVMSesqOp}
\hat{M}_f(U)=
\frac{1}{(2\pi\hbar)^4} \int_{\RNumb^{8}} d\rho(p,x) \,
\Ket{p,x} \chi(f(p,x) \in U) \Bra{p,x} \,.
\end{equation}
%

\section{Bounds of matrix elements} \label{appB}

Matrix elements of the observable $\hat{f}$ can be defined similarly to
expectation value \eqref{SpectralRepf}. If the required limit does not exist, all
matrix elements can be approximated by finite sums. In the case in which this limit
leads to an integral form, matrix elements of the observable $\hat{f}$ can be computed by
the following sesquilinear form:
\begin{equation}
\label{SesquiLinMf2}
\Bra{\psi_2} \hat{f} \Ket{\psi_1} \equiv \check{f}(\psi_2,\psi_1):=
\int_{\RNumb} u\, d\Bra{\psi_2} \hat{M}_f(u) \Ket{\psi_1} \, .
\end{equation}
This form is bounded by the $L^\infty$-norm of the corresponding quadratic form
$\check{f}(\psi,\psi)$, i.e., by the $L^\infty$-norm of appropriate expectation
values (see \cite{Mlak1972}):
\begin{equation}
\label{QuadrSesqBound}
\Norm{\Aver{\check{f}}}_{\infty}
\leq \sup\limits_{\Norm{\psi_1}=\Norm{\psi_2}=1}^{} |\check{f}(\psi_2,\psi_1)|
\leq 2\Norm{\Aver{\check{f}}}_{\infty} \,,
\end{equation}
where the $L^\infty$-norm is defined as $\Norm{\Aver{\check{f}}}_{\infty}
:=\sup\limits_{\Norm{\psi}=1}^{} \Aver{\check{f},\psi}$.

\section{Choice of the fiducial vector} \label{appC}

Let us try to choose $\Phi_0 $ in such a way that the integral
\begin{equation}\label{Ham16}
J := \int_{\RNumb^{4}}d^4 p\,g^{\alpha\beta}p_\alpha p_\beta
|\tilde{\Phi}_0 (p) |^2 \,
\end{equation}
vanishes. For this purpose, we use the fiducial vector in the form of the four-dimensional harmonic
oscillator ground state function
\begin{equation}\label{Ham17}
\Phi_0 (\xi) =  \prod_{\mu=0}^3
\left(\frac{\lambda_\mu}{\pi\hbar}\right)^{\frac{1}{4}}
\exp\left(-\frac{\lambda_\mu (\xi^\mu)^2}{2\hbar}\right)\,
\end{equation}
with $\lambda_0,\lambda_1=\lambda_2=\lambda_3 >0$, which is an even function of
$\xi^\mu$.
Its Fourier transform reads
\begin{equation}\label{Ham18}
\tilde{\Phi}_{0}(p) = \prod_{\mu=0}^3 (\pi \hbar \lambda_\mu)^{-1/4}
\exp\left(-\frac{(p_\mu)^2}{2 \hbar \lambda_\mu} \right) \, .
\end{equation}
Therefore, the expression for $J$ can be written as
\begin{eqnarray}
J & = & \int_{\RNumb^{4}}\,dp_0 dp_1 dp_2 dp_3\,
(- p_0^2 + p_1^2 + p_2^2 + p_3^2)\, (\pi \hbar \lambda_0)^{-1/2}
\exp\left(-\frac{p_0^2}{\hbar \lambda_0} \right) \nonumber \\
&& \times (\pi \hbar \lambda_1)^{-1/2}\exp\left(-\frac{p_1^2}{\hbar
  \lambda_1} \right)\,(\pi \hbar \lambda_2)^{-1/2}\exp\left(-\frac{p_2^2}{\hbar \lambda_2} \right) \nonumber \\
&& \times (\pi \hbar \lambda_3)^{-1/2}\exp\left(-\frac{p_3^2}{\hbar \lambda_3} \right)
\, .
\label{Ham19}
\end{eqnarray}
For further discussion we need the following  integrals
\begin{equation}\label{Ham20}
K_\mu := \int_{\RNumb}\,dp_\mu\, p_\mu^2
\exp \left( - \frac{p_\mu^2}{\hbar \lambda_\mu} \right)
= \frac{1}{2} \sqrt{\pi}(\hbar\lambda_\mu)^{\frac{3}{2}}, \quad \mu = 0,1,2,3
\end{equation}
and
\begin{equation}\label{Ham21}
L_\mu := \int_{\RNumb}\,dp_\mu\,
\exp \left( - \frac{p_\mu^2}{\hbar \lambda_\mu} \right)
=\sqrt{\pi\hbar\lambda_\mu}, \quad \mu = 0,1,2,3 \, .
\end{equation}
Using the above integrals, we get
\begin{equation}\label{ExplJ1}
J=\frac{1}{(\pi\hbar)^2\sqrt{\lambda_0\lambda_1\lambda_2\lambda_3}}
(-K_0L_1L_2L_3+L_0K_1L_2L_3+L_0L_1K_2L_3+L_0L_1L_2K_3)\, ,
\end{equation}
and, as $\lambda_1=\lambda_2=\lambda_3$, we obtain
\begin{equation}\label{ExplJ1B}
J=\frac{1}{(\pi\hbar)^2\sqrt{\lambda_0\lambda_3^3}}(-K_0L_3^3+3L_0K_3L_3^2)=
\frac{1}{2} \hbar \left(-\lambda_0+3\lambda_3\right)\, .
\end{equation}
Thus, $J=0$ if $\lambda_0=3\lambda_3$.



\end{document}